\documentclass[twocolumn,aps,prb,floatfix,showpacs]{revtex4}

\usepackage{graphics}

\begin{document}

\title{Influence of the Third Dimension of Quasi-Two-Dimensional
Cuprate Superconductors on Angle-Resolved Photoemission Spectra }

\author{A. Bansil$^1$, M. Lindroos$^{1,2}$, S. Sahrakorpi$^{1}$, and
R.S. Markiewicz$^{1}$}

\affiliation{ $^1$Physics Department, Northeastern University, Boston,
   Massachusetts 02115, USA \\
  $^2$Institute of Physics, Tampere University of Technology, P.O. Box
   692, 33101 Tampere, Finland}

\date{\today}

\begin{abstract}

   Angle-resolved photoemission spectroscopy (ARPES) presents
   significant simplications in analyzing strictly two-dimensional
   (2D) materials, but even the most anisotropic physical systems
   display some residual three-dimensionality. Here we demonstrate how
   this third dimension manifests itself in ARPES spectra of quasi-2D
   materials by considering the example of the cuprate
   Bi$_2$Sr$_2$CaCu$_2$O$_{8}$ (Bi2212). The intercell, interlayer
   hopping, which is responsible for $k_z$-dispersion of the bands, is
   found to induce an irreducible broadening to the ARPES lineshapes
   with a characteristic dependence on the in-plane momentum
   $k_\parallel$.  Our study suggests that ARPES lineshapes can
   provide a direct spectroscopic window for establishing the
   existence of coherent $c$-axis conductivity in a material via the
   detection of this new broadening mechanism, and bears on the
   understanding of 2D to 3D crossover and pseudogap and stripe
   physics in novel materials through ARPES experiments.

\end{abstract}

\maketitle

Angle-resolved photoemission spectroscopy (ARPES) has been applied
extensively in the recent years for investigating the electronic and
quasiparticle properties of high-temperature superconductors
\cite{damascelli03}. Much of the existing ARPES work on the cuprates
and other quasi-2D materials implicitly assumes perfect
two-dimensionality, ignoring the effects of dispersion in the third
dimension. As energy resolutions of the order of a few meV's have now
become possible in the state-of-the-art ARPES instrumentation, it is
natural to ask the question: How will $k_z$-dispersion play out in the
ARPES spectra of the quasi-2D materials? An obvious answer is that
spectral peaks will undergo shifts with photon energy --albeit small--
much like in the 3D case.\cite{inglesfield92,smith93,lindroos96} Our
analysis however reveals a very different scenario in that insofar as
ARPES response to $k_z$-dispersion is concerned, quasi-2D systems
differ fundamentally from their 3D counterparts. We find that in the
presence of typical final state dampings of the order of eV's, initial
state dispersions over much smaller energy scales will not appear as
energy shifts with $h\nu$, but will instead induce an irreducible
linewidth in the spectral peaks with a characteristic $k_\parallel$
dependence.

Our results bear on a variety of important issues in cuprate physics
and give insight into a number of puzzling features of ARPES in the
cuprates.  Since our new broadening mechanism does not have its origin
in a scattering process, it explains why the lifetimes derived from
ARPES spectra are generally found to be shorter than those obtained
from other experiments.  Moreover, Fermi surface (FS) maps obtained
via ARPES often display broad patches of spectral weight rather than
well-defined FS imprints, particularly near the antinodal point; such
patches occur naturally in our calculations.  Indeed, much of the {\it
pseudogap} phenomena -- in particular the lack of well-defined
quasiparticles near $(\pi ,0)$ -- will need to be reevaluated in light
of the present findings.  In this vein, these results also impact the
analysis of stripe physics through ARPES experiments. Finally,
detection of the $k_z$ related linewidths in the ARPES spectra offers
a unique spectroscopic window for establishing the existence of
coherent $c$-axis conductivity and intercell coupling in a system;
this important intrinsic property is not accessible directly through
other techniques.\cite{damascelli03}

Specifically, we focus in this article on the tetragonal body-centered
Bi$_2$Sr$_2$CaCu$_2$O$_{8}$ (Bi2212) compound, which has been a
workhorse of ARPES studies. Bi2212 is a nearly 2D material with two
CuO$_2$-layers in the primitive unit cell.  The {\em intracell}
interaction between the two CuO$_2$-layers, spaced a relatively short
distance of $\sim 3.2$~\AA\ apart, results in the well-known bilayer
splitting\cite{bansil99_2,lindroos02,feng01,chuang01,bogdanov01,damascelli03}.
On the other hand, the {\em intercell} coupling between the bilayer
slabs in different unit cells is expected to be smaller due to the
larger intercell Cu-Cu distance of $\sim 12$~\AA\, resulting in weak
but {\em non-vanishing} $k_z$-dispersion.  The intercell coupling will
be even more pronounced in other high-Tc's since Bi2212 presents one
of the longest $c$-axes in the cuprate family.

Concerning computational details, the band structure for Bi2212 was
obtained within the local-density-approximation (LDA) by using the
well-established Green's function methodology\cite{bansil99_1}. The
crystal potential used is the same as that employed in our previous
studies of Bi2212 and involves 30 atoms per conventional unit
cell\cite{bansil99_2,lindroos02,model}, yielding good agreement with
the experimentally observed Fermi surface
(FS)\cite{bansil02,asensio03,chuang03,bogdanov01,LDAscale}. ARPES
intensities have been computed within the one-step photoemission
formalism; see, Refs.~\onlinecite{bansil99_2}
and~\onlinecite{lindroos02} for details.

\begin{figure}
   \resizebox{6.5cm}{!}{\includegraphics{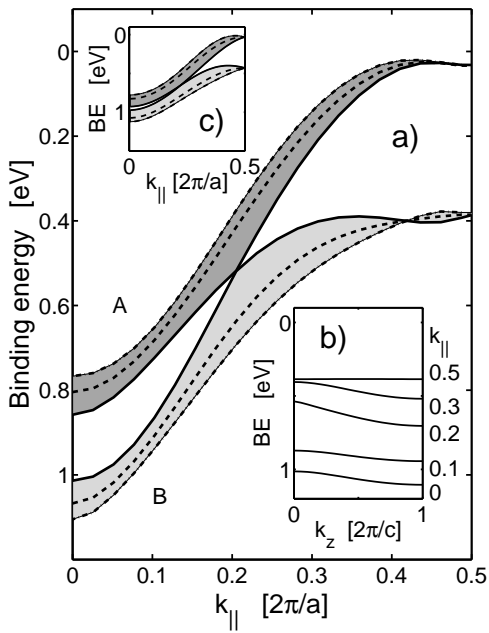}}
\caption{
   (a): Calculated first-principles ${\mathbf k}_\parallel$-dispersion
   in Bi2212 along [100]-direction for the B and A bands at three
   different $k_z$ values (in units of $2\pi/c$): $k_z=0$ (solid
   lines); $k_z=0.5$ (dashed), and $k_z=1$ (dash-dotted). Shading
   denotes the regions over which the bands wander as a function of
   $k_z$.  (b): $k_z$-dispersion of the B band at five different
   $k_{\parallel}$ values ranging from $0$ to $\pi/a$. (c): Same as
   (a), except that these results are based on a tight binding
   formalism.}
\label{fig:1}
\end{figure}
Figure~\ref{fig:1}(a), which considers the familiar antibonding (A)
and bonding (B) bands in Bi2212, shows that the ${\mathbf
k}_\parallel$-dispersion depends strongly on $k_z$ as the associated
bands wander over the two sets of shaded areas.  The $k_z$-dispersion
(i.e. the vertical width of the shaded areas) displays a striking
dependence on $k_\parallel$ and nearly vanishes at the antinodal point
$k_\parallel=(\pi/a,0)$. A clear bilayer splitting between the A and B
bands is seen at all $k$-values, except at $k_\parallel =
k^*_\parallel = 0.2(2\pi /a)$ for $k_z=0$ (solid lines), where A and B
bands touch. This level crossing leads to the anomalous
$k_z$-dispersion shown in Fig.~1(b) for the B band: The shape changes
from being described approximately by the form $\sin^2(k_zc/4)$ for
$k_\parallel = 0$, to $|\sin(k_zc/4)|$ at $k^*_\parallel$, and finally
back to $\sin^2(k_zc/4)$ for $k_\parallel a/(2\pi) = 0.5$. The
antibonding band behaves similarly (not shown for brevity). The
complex behavior of the bilayer splitting and the $k_z$-dispersion
shown in the first-principles computations of Figs.~1(a) and~1(b) can
be modeled reasonably well by a relatively simple tight-binding (TB)
Hamiltonian as seen from Fig.~1(c). We return below to discuss this
point further.

\begin{figure}
   \resizebox{8.5cm}{!}{\includegraphics{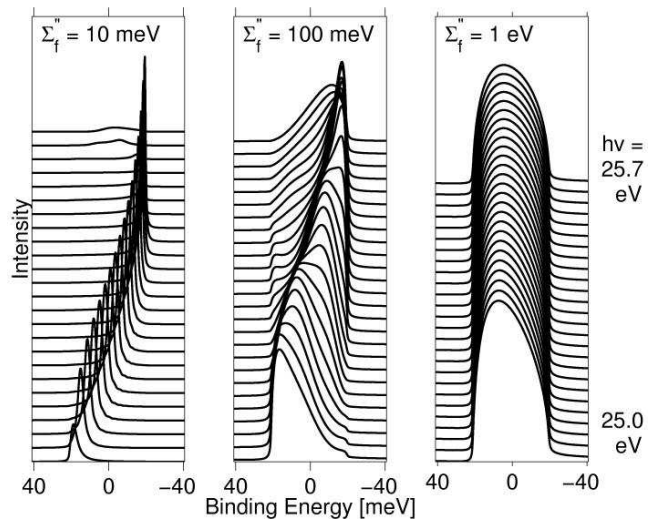}}
\caption{
   Simulated ARPES lineshapes (EDCs) in Bi2212 for a series of photon
   energies ($h\nu=25-25.7$~eV) at a {\em fixed}
   $k_\parallel=(0.34,0.09)2\pi/a$-point using three different values
   of the final state broadening given by the indicated imaginary parts
   of the self-energy, $\Sigma_f''$. In order to highlight the
   influence of $k_z$-dispersion, the initial state broadening is
   chosen to be very small, $\Sigma_i''=0.2$~meV.\cite{fermif} }
\label{fig:2}
\end{figure}
We next turn our attention to how $k_z$-dispersion will manifest
itself in ARPES spectra. The key insight in this regard is provided by
the simulations\cite{fermif} of Fig.~2, which consider the evolution
of the lineshape of the energy distribution curves (EDCs) as the final
state broadening introduced via the imaginary part of the final state
self-energy\cite{lifetime}, $\Sigma_f''$, is increased from a very
small value in (a) to a realistic value in (c). In (a), the position
of the spectral peak undergoes the familiar shift as $h\nu$ varies.
This shift results from the fact that in the photo-excitation process
$k_\parallel$ must remain unchanged in transmission of the electron
across the surface and the initial and final states can connect only
at a specific value of $k_z$ in order to conserve energy. Note that
the total shift in the peak position in the EDCs of (a) gives the size
of the $k_z$-dispersion of the initial state. This should not be
confused with the change in $h\nu$ needed to probe such a band. The
change in $h\nu$ is controlled by the final state dispersion, which is
generally much larger than that of the initial state.

Continuing to the intermediate case of $\Sigma_f''= 0.1$~eV in
Fig.~2(b), the shift in the peak position from the bottom to the top
of the initial state band is once again evident, but the lineshapes
are quite different, even though the initial state damping
$\Sigma_i''$ in (b) is identical to that in (a). It is striking that
some spectral intensity appears in (b) at all energies encompassed by
the initial state band at every $h\nu$. This remarkable effect comes
about because the energy uncertainty permitted by the width of the
final state allows the photoelectron to couple with initial states
off-the-energy-shell. The lineshape thus develops a new component with
an $h\nu$-independent width equal to the initial state bandwidth,
which rides on top of the energy conserving peak in (a).  Notice also
how the changes in the peak position could allow observing different
FSs as one maps different values of $k_z$. Finally, for the realistic
final state width of $\Sigma_f''=1$~eV in (c), the initial state
bandwidth component dominates the lineshape. The lineshape of the EDC
curve is now virtually $h\nu$-independent. Despite the large
final-state broadening, the energy spread of the EDC remains equal to
the {\em initial state bandwidth in $k_z$-direction} because outside
of this interval, there are no initial state electrons capable of
absorbing the photon.

\begin{figure}
   \resizebox{6.5cm}{!}{\includegraphics{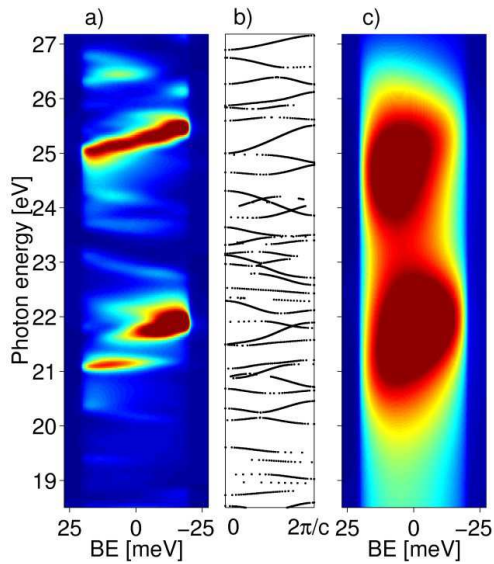}}
\caption{
   (Color) (a) Simulated ARPES intensities of Fig.~2(b) are depicted
   over a wide photon energy range of $19-27$~eV for $\Sigma_f''=
   0.1$~eV. Reds denote highs. (b) Final state band structure as a
   function of $k_z$. (c) Same as (a) except that $\Sigma_f''=1 $~eV,
   which corresponds to the case of Fig.~2(c).  }
\label{fig:3}
\end{figure}
Fig.~3 elaborates on these points by considering some of the results
of Fig.~2 over a much broader range of photon energies. Fig.~3(a)
presents the spectra of Fig.~2(b) for $\Sigma_f''= 0.1$~eV for $h\nu
\approx 19-27$~eV.  Different colored bands here represent the same
initial states being excited to different final states. Comparing
Figs.~3(a) and~3(b), we see that the peaks of Fig.~3(a) more or less
follow the final state band structure of Fig.~3(b) [The agreement is
not expected to be perfect of course, since the initial state
dispersion is not entirely negligible]. In Fig.~3(c), when a more
realistic final state width of $\Sigma_{f}''=1$~eV is assumed, much of
the structure is lost. The large variations in colors (or intensities)
observed in Figs.~3(a) and~3(c) are the consequence of the well-known
$k_\parallel$ and $h\nu$ -dependency of the ARPES matrix
element.\cite{bansil99_2,lindroos02,asensio03,sahrakorpi03,damascelli03}

The preceding discussion of Figs.~2 and~3 makes it obvious that in a
quasi-2D system, $k_z$-dispersion leads to an irreducible linewidth in
the ARPES spectra, which cannot be resolved by changing photon
frequency. This effect will also appear in the FS maps observed by
measuring emission from $E_F$ in the $(k_x,k_y)$-plane. Fig.~4
illustrates how this plays out.  The standard bilayer split FS in the
$k_z=0$ plane, given in Fig. 4(a), is the FS usually thought to be
measured in ARPES.  In Fig.~4(b) the full 3D FS has been projected on
to the (001)-plane by collecting individual FS cuts corresponding to
different $k_z$ values. The $k_z$-dispersion is now seen to introduce
an effective "broadening" to the FS imprint. In a perfect 2D-system,
the maps in (b) and (a) would be identical. The effect of using a
finite energy window of $\pm 30$~meV around $E_F$ in the computations
is shown in Figs.~4(c) and~4(d). These simulations indicate the
influence of a finite experimental resolution on the
results.\cite{LDAscale,window} Note that even with the window of $\pm
30$~meV in (d), the broadening effect of $k_z$-dispersion is not
washed out. In fact, the broadening is somewhat enhanced, especially
near the antinodal point due to the contribution of the flat bands
related to the van Hove singularity (VHS).
\begin{figure}
   \resizebox{6.5cm}{!}{\rotatebox{00}{\includegraphics{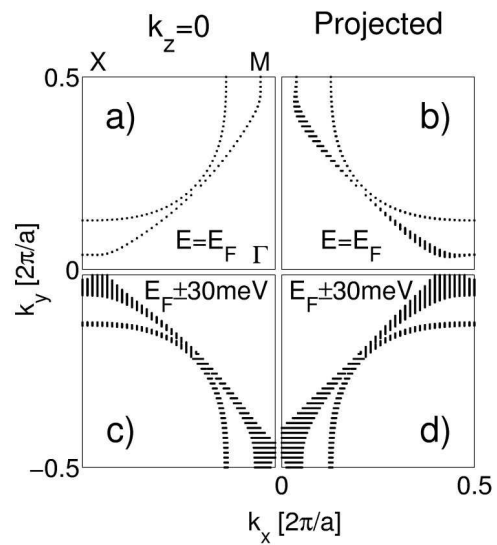}}}
\caption{
   Different imprints of the FS of Bi2212. (a): Section of the FS at
   $k_z=0$.  (b): Projection of the FS on to the (001)-plane, where
   the "broadening" of the FS reflects the effect of
   $k_z$-dispersion. (c) and (d) correspond to (a) and (b),
   respectively, except that the effect of adding a finite energy
   window of $\pm 30$~meV is included in the computations. }
\label{fig:4}
\end{figure}

It is also interesting to consider changes in the linewidth in
momentum, $\Delta k$, as one moves away from the Fermi level. In
general, our simulations indicate that $\Delta k$ increases with
increasing binding energy (BE), due mainly to the flattening of bands
and the concomitant reduction in the band velocity.\cite{smith93} In
any event, the broadening as a function of BE is neither simply
quadratic nor exponential.  Furthermore, considering the the effect of
varying $\Sigma_{i}''$ at a general $k_\parallel$-point, we have found
that as the initial state damping due to intrinsic scattering
mechanisms (simulated via the value of $\Sigma_{i}''$) decreases, the
linewidth $\Delta E$ becomes increasingly dominated by the irreducible
width associated with $k_z$-dispersion.  Along the antinodal direction
this dispersion is negligible, and $\Delta E$ and $\Sigma_i''$ are
related linearly.\cite{splitting}

We return now to comment briefly on the bands of Figs.~1(a) and~1(b).
Insight into how the complex $k_\parallel$- and $k_z$-dependencies of
these bands reflect {\em intercell} as well as {\em intracell} hopping
effects can be gained by modeling these bands within the TB
framework. In the absence of intercell coupling, the conventional
bilayer splitting possesses the form $ t_{bi}=t_z(c_x-c_y)^2$, with
$c_i=cos(k_ia)$, $i=x,y$. The {\em intercell} coupling in the cuprates
may be included in the one-band model with dispersion\cite{OKA}
\mbox{$ \epsilon_k = -2t(c_x+c_y)-4t'c_xc_y - T_z( k_{\parallel},s_z)
[{(c_x-c_y)}^2/4+a_0]$,} where $s_z=sin(k_zc/4)$ and $a_0=0.6$
corrects for the finite bilayer splitting found at $k_\parallel =0$.
The form of $T_z$ depends on the particular cuprate considered.  While
Ref.~\onlinecite{OKA} lists seven inequivalent interlayer hopping
parameters, we find that we can describe the dispersion reasonably by
including only the dominant contribution associated with hopping
between Cu $4s$ levels by introducing $ T_z =
\pm\sqrt{(t_z-t'_z)^2+4t_zt'_zs_z^2}$, where plus (minus) sign refers
to the bonding (antibonding) solution.  $t_z$ is a constant associated
with intracell interlayer hoppings, and $
t'_z=4t'_{z0}cos{(k_xa/2)}cos{(k_ya/2)}$ accounts for intercell
hopping. The extra angular dependence in this term arises because for
intercell hopping the two CuO$_2$-planes are offset, so that one Cu
atom sits above an empty cell.  Note that for $t'_z=0$ one obtains the
simple bilayer splitting $t_{bi}$ with no $k_z$-dispersion.  TB bands
of Fig.~\ref{fig:1}(c) assume (in eV): $t=0.42$, $t'=-0.12$,
$t_z=0.125$, and $t'_{z0}=0.04$. A comparison of the first principles
and TB bands in Figs.~\ref{fig:1}(a) and~\ref{fig:1}(c), respectively,
shows that our TB model reproduces the bilayer splitting, the collapse
of the $k_z$-dispersion around $(\pi/a,0)$, and the anomalous
dispersions associated with level crossing to a reasonably good
degree, some minor discrepancies notwithstanding.

We emphasize that $t_z'$, which provides intercell coupling, is
essential for obtaining coherent $c$-axis conductivity. $t_z'$ thus
controls the intrinsic resistivity anisotropy and gives a measure for
discriminating between coherent and incoherent $c$-axis
hopping.\cite{anomaloustemp} The lifetimes in the cuprates extracted
from ARPES are considerably smaller than optical lifetimes or those
deduced from transport or tunneling measurements\cite{tunn}. A part of
this difference may be explained to be the result of an unresolved
bilayer splitting\cite{BoLa}. The remaining anomalous broadening is
often interpreted\cite{AVar} in terms of small-angle scattering, which
does not contribute to transport, but its origin remains
obscure\cite{MilDrew}. The present analysis indicates that part of the
broadening of ARPES lines has its origin in the $k_z$-dispersion,
unrelated to any scattering mechanism.

Our predicted effects of $k_z$-dispersion on the linewidths in Bi2212
should be resolvable with the power of currently available high
resolution ARPES instrumentation.\cite{renormalization} The size of
our novel line broadening mechanism will be larger in materials with
greater 3D character (e.g. YBCO, LSCO, NCCO, ruthnates, manganites,
etc.), although we expect the results of Figs.~1-4 to provide at least
a qualitative handle in quasi-2D materials more generally. Notably,
coherent 3D coupling has recently been reported in a Tl
cuprate\cite{Hus}. Additionally, a similar $k_z$-induced broadening
may be expected to have an impact on other experimental probes, such
as Raman scattering\cite{DeKa}.

In conclusion, we have demonstrated that residual $k_z$-dispersion in
quasi-2D materials will induce an {\em irreducible linewidth} in ARPES
spectra.  This intrinsic linewidth offers a new spectroscopic window
for understanding in- as well as out-of-plane scattering mechanisms
and the nature of the 2D to 3D crossover in the cuprates. This highly
anisotropic line broadening mechanism, which has not been recognized
previously and is unrelated to 2D physics, indicates that the existing
analysis of stripe and pseudogap physics based on ARPES spectra should
be reexamined. Our study shows how ARPES can be extended to unravel
the hidden third dimension of energy bands and Fermi surfaces in
quasi-2D systems with wide-ranging consequences for understanding the
nature of electronic states in many novel materials.

\vspace*{0.5cm}
\begin{acknowledgments}

   This work is supported by the US Department of Energy contract
   DE-AC03-76SF00098, and benefited from the allocation of
   supercomputer time at NERSC, Northeastern University's Advanced
   Scientific Computation Center (ASCC), and the Institute of Advanced
   Computing (IAC), Tampere.  One of us (S.S.) acknowledges Suomen
   Akatemia and Vilho, Yrj\"o ja Kalle V\"ais\"al\"an Rahasto for
   financial support.

\end{acknowledgments}


\begin{thebibliography}{99}
\bibitem{cdef}
   Throughout this article, $z$-axis is defined to be normal to the
   Cu-O planes.
\bibitem{model}
   The well-known ``Bi pockets'' in the vicinity of the $M$-point have
   been lifted above the Fermi energy by modifying the Bi and O
   potentials, consistent with experimental data.
\bibitem{LDAscale}
   It should be kept in mind that LDA calculations systematically
   overestimate band dispersions, typically by a factor of about 2-4.
\bibitem{fermif}
   The spectra have not been convoluted with the Fermi function
   throughout this article.
\bibitem{lifetime}
   The imaginary part of the self-energy $\Sigma''$ is proportional to the
   inverse of the lifetime of the (quasi)particle in that state.
\bibitem{window}
   The $\pm 30$~meV window used would simulate an effective
   experimental resolution of about $\pm 8$~meV [see
   Ref.~\onlinecite{LDAscale}].
\bibitem{anomaloustemp}
   Of course, in Bi2212, the anomalous temperature dependence of
   $\rho_c$ strongly hints that it is incoherent.
\bibitem{renormalization}
   Note from Fig.~1 that a typical broadening is a significant fraction
   of the full bilayer splitting, which is experimentally found to be
   about 100~meV [recall Ref.~\onlinecite{LDAscale}].
\end{thebibliography}
\end{document}